\documentclass[12pt]{elsarticle}
\usepackage{amssymb,amsmath}
\usepackage[english]{babel}    
\usepackage[latin1]{inputenc}   
\usepackage{subfigure}
\usepackage{picinpar}
\usepackage{color}
\usepackage{lineno}

\newcommand{\ie}{{\it i.e.}}

\newcommand{\eg}{{\it e.g.}}

\begin{document}

\begin{frontmatter}

\title{Multi-$q$ Analysis of Image Patterns}

\author[iprj]{RICARDO FABBRI}
\ead{rfabbri@gmail.com}

\author[ifsc]{WESLEY N. GON\c{C}ALVES}
\ead{wnunes@ursa.ifsc.usp.br}

\author[ufrj]{FRANCISCO J. P. LOPES}
\ead{fjplopes@gmail.com}

\author[ifsc]{ODEMIR M. BRUNO}
\ead{bruno@ifsc.usp.br}

\address[ifsc]{Instituto de F\'{i}sica de S\~{a}o Carlos (IFSC), Universidade de S\~{a}o Paulo (USP)\\ Av.  Trabalhador S\~{a}o Carlense, 400\\
13560-970 - S\~{a}o Carlos, SP, Brasil\\phone/fax: +55 16 3373 8728 / +55 16 3373 9879}

\address[iprj]{Instituto Polit\'{e}cnico do Rio de Janeiro, Universidade do Estado do Rio de Janeiro\\Caixa Postal: 97282 -  CEP 28601-970 - Nova Friburgo, RJ, Brasil}

\address[ufrj]{Instituto de Biof\'{i}sica Carlos Chagas Filho, Universidade Federal do Rio de Janeiro\\Bloco G - 1 andar - Sala G1-019 - Cidade Universit\'{a}ria\\ 21941-902 - Rio de Janeiro, RJ, Brasil}

\begin{abstract}
This paper studies the use of the Tsallis Entropy versus the classic
Boltzmann-Gibbs-Shannon entropy for classifying image patterns. Given a database
of 40 pattern classes, the goal is to determine the class of a given image
sample.  Our experiments
show that the Tsallis entropy encoded in a feature vector for different $q$
indices has great advantage over the Boltzmann-Gibbs-Shannon entropy for
pattern classification, boosting recognition rates by a factor of 3.
We discuss the reasons behind this success, shedding light on the usefulness of
the Tsallis entropy.
\end{abstract}


\end{frontmatter}

\section{Introduction}

Image pattern classification is an important problem for a number of fields, \eg,
for recognizing embryo development stages~\cite{Zhong:etal:2009}, determining protein
profile from flourescence microscopy images~\cite{Janssens:etal:DGE2005}, and identifying
cellular structures~\cite{Bohm:etal:PNAS2000}. This can be a challenging problem, 
specially for a large number of classes, with many different solutions proposed in
the literature~\cite{Duda:etal:2001}. In this paper, however, our
primary goal is to study the usefulness of the Tsallis
entropy~\cite{Tsallis:book:2009} by comparing it to the classic
Boltzmann-Gibbs-Shannon (BGS) entropy when applied to classifying image
patterns.

In statistical mechanics, the concept of entropy is related to the distribution
of states, which can be characterized by the system energy levels. From an
information-theoretic point of view, entropy is related to the lack of information
of a system. It also represents how close a given probability distribution is to
the uniform distribution, \ie, it is a measure of randomness, peaking at
the uniform distribution itself.  For image pattern classification, this interpretation
can be useful since, for example, a symmetric, periodic or smooth image has less
``possible states'' than more uniformly random images. A direct link between
probability distributions and such concept of entropy was proposed
by Boltzmann and is the foundation of what is now known as the
Boltzmann-Gibbs-Shannon (BGS) entropy. A
well-know generalization of this concept, the Tsallis
entropy~\cite{Tsallis:book:2009}, extends its application to
so-called non-extensive systems using a new parameter $q$. It recovers the
classic entropy for $q \to 1$, and is better suited for long-range
interactions between states (\eg, in large pixel neighborhoods) and long-term memories. The Tsallis generalization of entropy has a vast
spectrum of application, ranging from physics and chemistry to computer science. For
instance, using the non-extensive entropy instead of the BGS entropy can produce gains in
the results and efficiency of optimization
algorithms~\cite{Tsallis:Stariolo:PhysicaA1996}, image
segmentation~\cite{Chang:etal:ISPV2006,Albuquerque:etal:PRL2004,Rodrigues:Giraldi:SIBGRAPI2009} or edge detection
algorithms~\cite{Hamza:etal:JEI2006}.

In this paper we study the power of the Tsallis entropy in comparison to classic
BGS entropy for the construction of feature vectors for classifying image
patterns or textures. Given a database of patterns, typically comprising 40
classes, the goal is to determine the class of a given image sample. Our
experiments show that the Tsallis entropy encoded in a feature vector for
different $q$ indices produces great advantage over the Boltzmann-Gibbs-Shannon
entropy for this problem, boosting recognition rates by a factor of 3.

This paper is organized as follows. A review of definitions and notation of
fundamental concepts is given in Section~\ref{sec:notation}. Our is described
Details about the problem of
image pattern classification and our approach based on the Tsallis entropy are described in
Section~\ref{sec:multiq}. The basic experimental setup is provided in
Section~\ref{sec:experimental:setup}. In Section~\ref{sec:results} 
experimental results towards analyzing the power of the Tsallis entropy for image pattern
classification are described. Section~\ref{sec:conclusion} revises and concludes the paper.


\section{Formulation and Notation}\label{sec:notation}

Assume ${p_i}$ is the probability distribution or the histogram of graylevels $g_i=1,\cdots,W$ in the
grayscale image $I$, \ie, $p_i$ equals the number of pixels having intensity $i$ divided by the
total number of pixels. We assume $g_1 = 0$ stands for black, and $g_{256}=255$ for white.
The number $W$ of different graylevels is typically $256$ for 8-bit images. The
BGS entropy is defined as:\footnote{We have dropped a constant of direct proportionality for the purpose of this paper.}
\begin{equation}
S_{BG} = - \sum_{i=1}^{W} p_i \log{p_i}, \qquad \sum p_i = 1.
\end{equation}
In the special case of a uniform distribution, $p_i = 1/W$, so that $S_{BG} = \log W$.
Similarly, the Tsallis entropy is defined as
\begin{equation}
S_{q} = \frac{1 - \sum p_i^q}{q-1},
\end{equation}
which recovers BGS entropy in the limit for $q\to 1$. The relation to BGS
entropy is made clearer by rewriting this definition in the form:
\begin{equation}
S_q(p) = -\sum_i p_i^q \ln_q p_i,
\end{equation}
where 
\begin{equation}
\ln_q(x) \dot= \frac{x^{1-q} - 1}{1 - q}
\end{equation}
is called the $q$-logarithm, with $\ln_q(x)\to\ln x$ for $q\to 1$. For any value
of $q > 0$, $S_q$ satisfies similar properties to the BGS entropy; for instance,
$S_q \geq 0$, and $S_q$ attains its maximum at the uniform distribution.

The BGS entropy is additive in the sense that the entropy of the whole system
(the entropy of the sum) coincides with the sum of the entropies of the parts.
This is not the case for the Tsallis entropy when $q\neq 1$, however.
Formally,
\begin{align}
&S_{BG}(A+B) = S_{BG}(A) + S_{BG}(B),\\
\intertext{while}
&S_q(A+B) = S_q(A) + S_q(B) + (1-q)S_q(A)S_q(B).
\end{align}

\section{Multi-q analysis}\label{sec:multiq}

\begin{figure}[!h]
\centering
\includegraphics[width=1\textwidth]{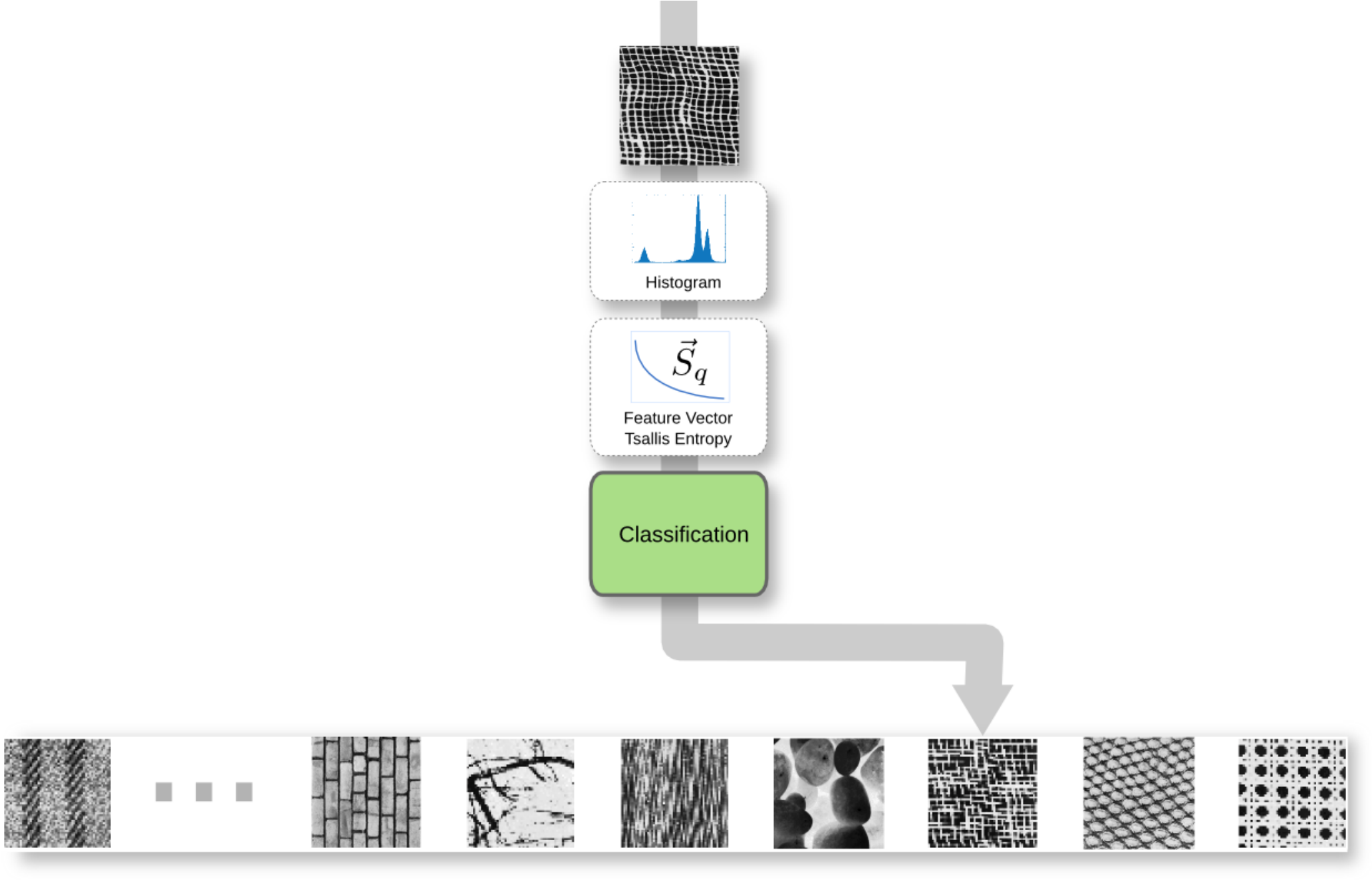}%
\caption{%
Our approach to image pattern classification works by extracting the histogram
of an input image sample, computing a vector of Tsallis Entropies, and using
that in a classifier to produce a class label among 40 classes stored in a
database. The image samples in this figure are real samples illustrating the
Brodatz database used in this paper.
}\label{fig:system:drawing}
\end{figure}

The primary goal of image pattern classification is to assign a class label to a
given image sample or window, the label being chosen among a predefined set of
classes in a database.  
Figure~\ref{fig:system:drawing} shows a schematic of the classification process.

In supervised classification, the classifier is trained from a set of samples that
are known to belong to the classes (\emph{a priori} knowledge). The classifier is then
validated by another set of samples.  This methodology can be used for
pattern recognition tasks as well as for mathematical modeling.  
Traditionally,
in image analysis, a feature vector is extracted from the image and used to
train and validate the classifier.  It is expected that the feature vector
concentrates the most important information about the image.   


In this work we investigate the Tsallis entropy as tool to analyze image
information and compare it to the traditional BGS entropy. Beyond
statistical mechanics, the BGS entropy is also
traditionally used in information theory, and is present as a metric in many
image analysis methods, for instance Gabor texture analysis, Fourier analysis,
wavelet, shape analyses among many others.
A classical and simple problem in image analysis is considering
the distribution of pixel intensities in an image as a measure of texture, by analyzing
its histogram. Such an approach has been used since the 70's and, despite its
simplicity, provides good results and is still subject of active
research~\cite{Barbieri:etal:PhysicaA2011}. Therefore, we
have decided to use histogram texture analysis to investigate the potentiality of Tsallis
entropy applied to information theory in the context of images, and comparing its results to the
those obtained with BGS entropy alone.

Histogram texture analysis begins by computing the image histogram ${p_i}$
of intensities, where $p_i$ is the number of pixels in the image for each
intensity $i$.  Assuming 8-bit grayscale images, it abstracts the image
information into a feature vector of 256 dimensions.  The histogram encodes a
mixture of multiple intensity distributions representing luminosity patterns of
image subsets, therefore being a clear candidate for image pattern
representation for a number of classification applications.  Although the
histogram is largely used in image analysis, it is
limited, due to its simplicity.  For instance, the spatial information is not preserved by the
histogram.  Different images that have the same distribution of pixels have the
same histogram; for instance, consider two images: a checkerboard pattern and an image
split in the middle in black and white.  While the visual information presented
is quite different, they have the same histogram.

Despite its limitations, the image histogram has been used for different
purposes, achieving good results, e.g., in image segmentation, image thresholding and
pattern recognition.  In this work, we are taking into account the third
alternative applied to image classification.  To classify an image or an image
sample based on the histogram, statistical metrics are traditionally employed,
such as mean, mode, kurtosis and BGS entropy.  Therefore, the simplicity
and popularity of the image histogram can help focus the results of the
classification on the entropy analysis itself.
  


The concept of Tsallis entropy (and, in particular, BGS entropy) defined
in Section~\ref{sec:notation} provides ways to further abstract the information of the
intensity histogram.  For $q=1$ we have the classic BGS entropy, strongly
abstracting the 256-dimensional histogram into the extreme case of a single
number $S_1$. This
paper explores multiple $q$ parameters towards forming better feature vectors
for classification. We construct feature vectors of the form $\vec S_q =
(S_{q_1},\dots,S_{q_n})$, Figure~\ref{fig:system:drawing}, whose dimension $n$
(typically 4--20) provides a middle ground between the total abstraction of 1D
BGS entropy and the full 256-dimensional histogram. The experiments show that very few
dimensions of Tsallis-entropy values in $\vec S_q$ are already enough to
outperform BGS entropy by a large factor.

\section{Experimental Setup}\label{sec:experimental:setup}

The database used to evaluate our approach was created from
Brodatz's art book~\cite{Brodatz:book:1966}.
This book is a black and white photography study for art and
design and it was carried out on different patterns from wood, grass, fabric, among
others. The Brodatz database became popular in the imaging sciences and is widely
used as a benchmark for the visual attribute of texture.  The database used in
our work consists of 40 classes of texture, where each class is represented by a
prototypical photograph of the texture containing no other patterns.  Such
$512\times512$ images are scans of glossy prints that were acquired from the
author.  A given image sample to be classified is much smaller than the class
prototype image, $200\times200$ in this paper.  The $512\times512$ prototype
image can generate numerous $200\times200$ image samples representing the same
class using a sliding window scheme.  To construct our final database we perform
this sliding window process to extract 10 representative image samples for each
class.  Therefore, any incoming sample to be classified is of the same size as
the training windows. A few samples from this database are ilustrated in
Figure~\ref{fig:system:drawing}.

Several approaches for the task of classification have been proposed in the
literature.  Since our focus is on image representation, we used the simple and
well-known Naive Bayes classifier rather than a more sophisticated classifier.
Although more sophisticated classifiers have been shown to produce superior
results (e.g.  multilayer perceptron and support vector machine), we are
interested in showing the discrimination power provided by the features
themselves rather than showing the classification power of classifiers.

In order to objectively evaluate the performance of the Tsallis entropy versus
BGS entropy for classification, we use the stratified 10-fold cross-validation
scheme~\cite{witten2011data}.  In this scheme, the samples are randomly divided
into 10 folds, considering that each fold contains the same proportions of the classes 
(i.e. for the Brodatz dataset, each fold contains 40 samples, one sample of each class).
At each run of this scheme, the classifier is trained using all but one fold and then 
evaluated on how it classifies the samples from the separated fold.  This process is repeated such 
that each fold is used once as validation.  The performance is averaged,
generating a single number for classification rate which represents the overall
proportion of success over all runs.  A standard deviation is also computed and
displayed when significant.  In the next section, moreover, a confusion matrix
is eventually generated to analyze the performance of a specific classifier strategy.
The confusion matrix is very well known in statistical classification and
artificial intelligence.  It is an $n\times n$ matrix where $n$ is the number of
classes, and whose entry $(i,j)$ expresses how many patterns of class $i$ were
labeled as class $j$.  It allows analyzing the error of the classification and
which class was most wrongly classified.

\section{Experimental Results}\label{sec:results}

We have conducted two sets of experiments to evaluate the texture recognition performance.
The aim of the first set of experiments is to analyze the power of the Tsallis entropy with
only one $q$ value and compare its performance against that of the BGS entropy.
The second set of experiments is devised to analyze the Tsallis entropy with a
multi-$q$ approach, including analysis of different sets of $q$'s.

\subsection{Classification Results: single $q$}

\begin{figure}
\centering
\scalebox{0.4}{\includegraphics{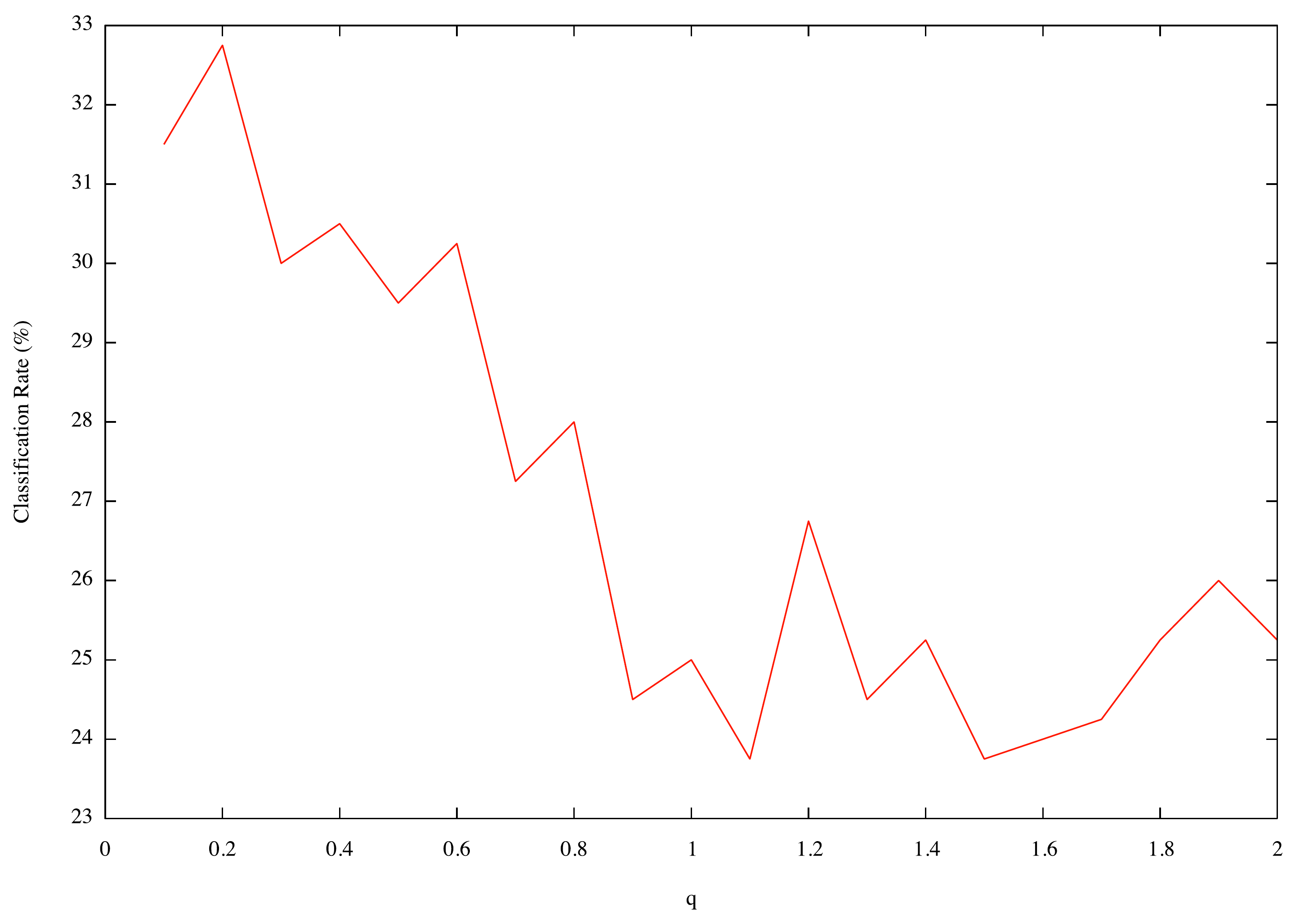}}%
\caption{%
Classification rate using a single value of $q$ at a time.
}\label{fig:classific:versus:q}
\end{figure}

To fairly compare the BGS and Tsallis entropies, we conducted an experiment
using a single value of $q$, that is, the feature vector reduced to a single
number for the purpose of our study. Note, however, that in a practical system one would
use a higher-dimensional feature vector (i.e., more numbers) to represent an
image sample. In the Brodatz dataset, using the BGS entropy alone
yielded a classification rate of $25\% (\pm 5.78)$.
To compare it to the Tsallis entropy, we need to choose an appropriate value of $q$.
The classification rate of different values of $q$ are presented in the plot of the Figure~\ref{fig:classific:versus:q}.
The best result of $32.75\%(\pm 6.17)$ was obtained by the Tsallis entropy with $q=0.2$.
Notice that any Tsallis entropy with $q < 1$ outperforms the BGS entropy. 
Moreover, note that in the most cases, the classification rate decreases as the value of $q$ increases.

We expected that for texture images in general (i.e., beyond Brodatz) the
highest classification rates are also obtained for values of $q$ close
to $0.2$.  Table~\ref{tab:datasets} presents the best values of $q$ for
different texture image datasets, including CUReT~\cite{Dana:etal:ACMTOG1999}
(Columbia-Utrecht Reflectance and Texture Database), Outex from the University of
Oulu~\cite{Ojala:etal:ICPR2002}, and VisTex from MIT.  For all but the VisTex dataset the
best value of $q$ was $0.2$.  For VisTex the highest classification
rate of $21.30\%(\pm 3.27)$ was obtained by $q=0.1$ while $q=0.2$ achieved a
correct classification rate of $20.72\%(\pm 3.21)$.  Moreover, the Tsallis entropy
using $q=0.2$ outperforms the BGS entropy for all datasets. Note that these
results concern the effectiveness of a \emph{single number} to abstract the information
of an image window. We strees that on a practical system one would likely represent a
$200\times200$ image with a larger set of numbers (features).

\begin{table}[h]
\centering
\begin{tabular}{|l|c|c|c|}
\hline
\textbf{Dataset} & \textbf{best q} & \textbf{Classification Rate} & \textbf{BGS Entropy} \\\hline
\hline
Brodatz & 0.2 & $32.75\%(\pm 6.17)$ & $25.00\%(\pm 5.78)$ \\\hline
VisTex &  0.1 & $21.30\%(\pm 3.27)$ & $16.90\%(\pm 3.26)$ \\\hline
CUReT & 0.2 & $10.14\%(\pm 0.84)$ & $08.64\%(\pm 0.77)$ \\\hline
Outex & 0.2 & $14.19\%(\pm 2.37)$ & $13.38\%(\pm 2.03)$ \\\hline
\end{tabular}
\caption{Classification results for different texture image datasets  and the
best value of $q$. The Tsallis entropy outperforms the BGS entropy for all datasets.}
\label{tab:datasets}
\end{table}

\subsection{Classification Results: multiple $q$}
\begin{figure}[!h]
\centering
\includegraphics[width=0.9\textwidth]{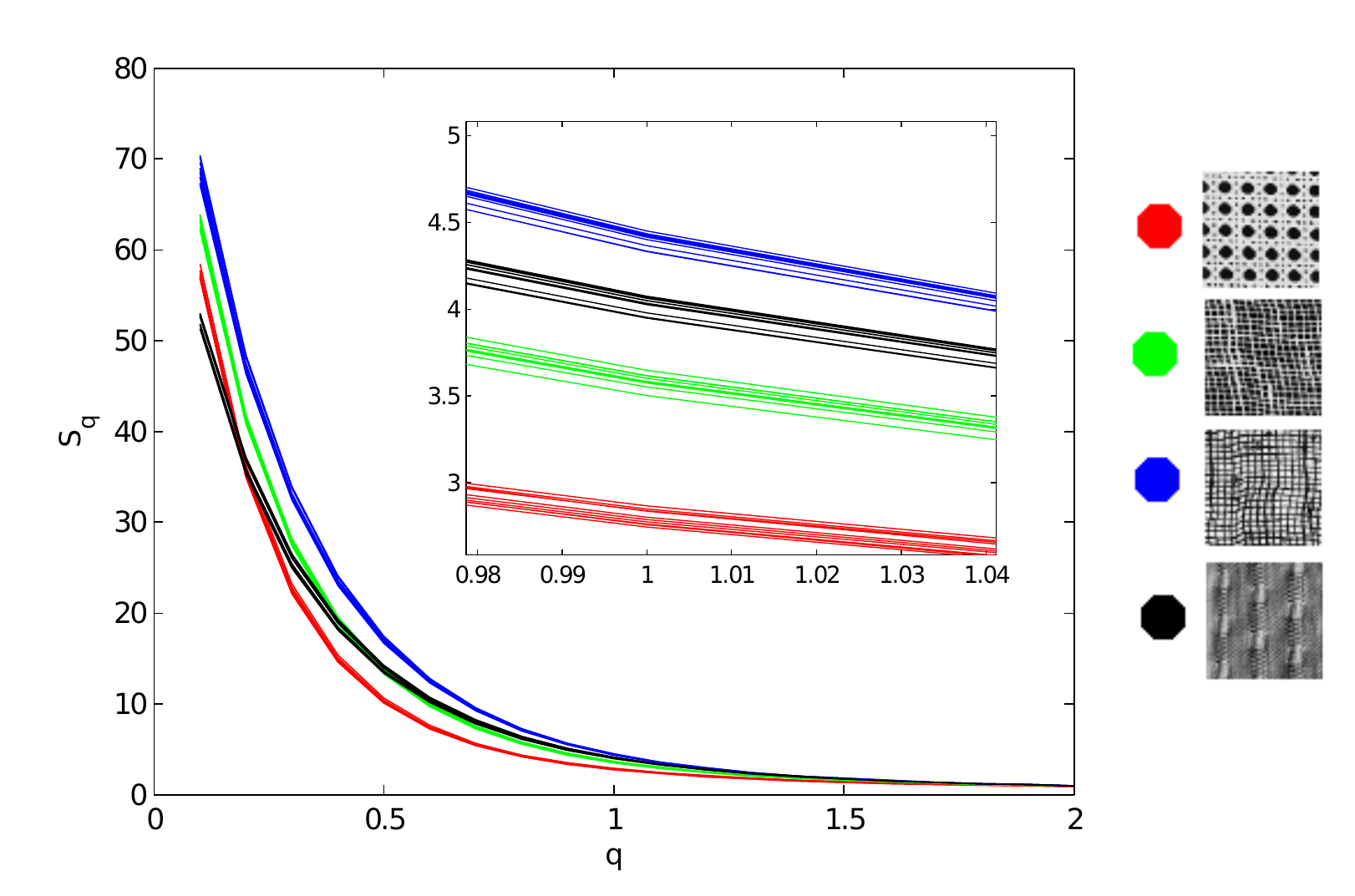}%
\caption{%
The feature vector $(S_{0.1}, ..., S_2)$ for three different textures, giving an idea of the discriminating power of the Tsallis entropy.
}\label{fig:feature:vectors}
\end{figure}

One hypothesis for the power of the Tsallis entropy concerning pattern
recognition is that each $S_q$ could hold different information about the pattern.
Therefore, different values of $q$ used together could improve classification
rates. Indeed, Figure~\ref{fig:feature:vectors} shows that the $S_q$ curve can help in
distinguishing the patterns --
the feature vector $\vec S_q = (S_{0.1},S_{0.2} ..., S_2)$ is plotted
for three different textures, giving an idea of the discriminating
power of the Tsallis entropy.

\begin{figure}[!hp]
\centering
\includegraphics[width=0.95\textwidth]{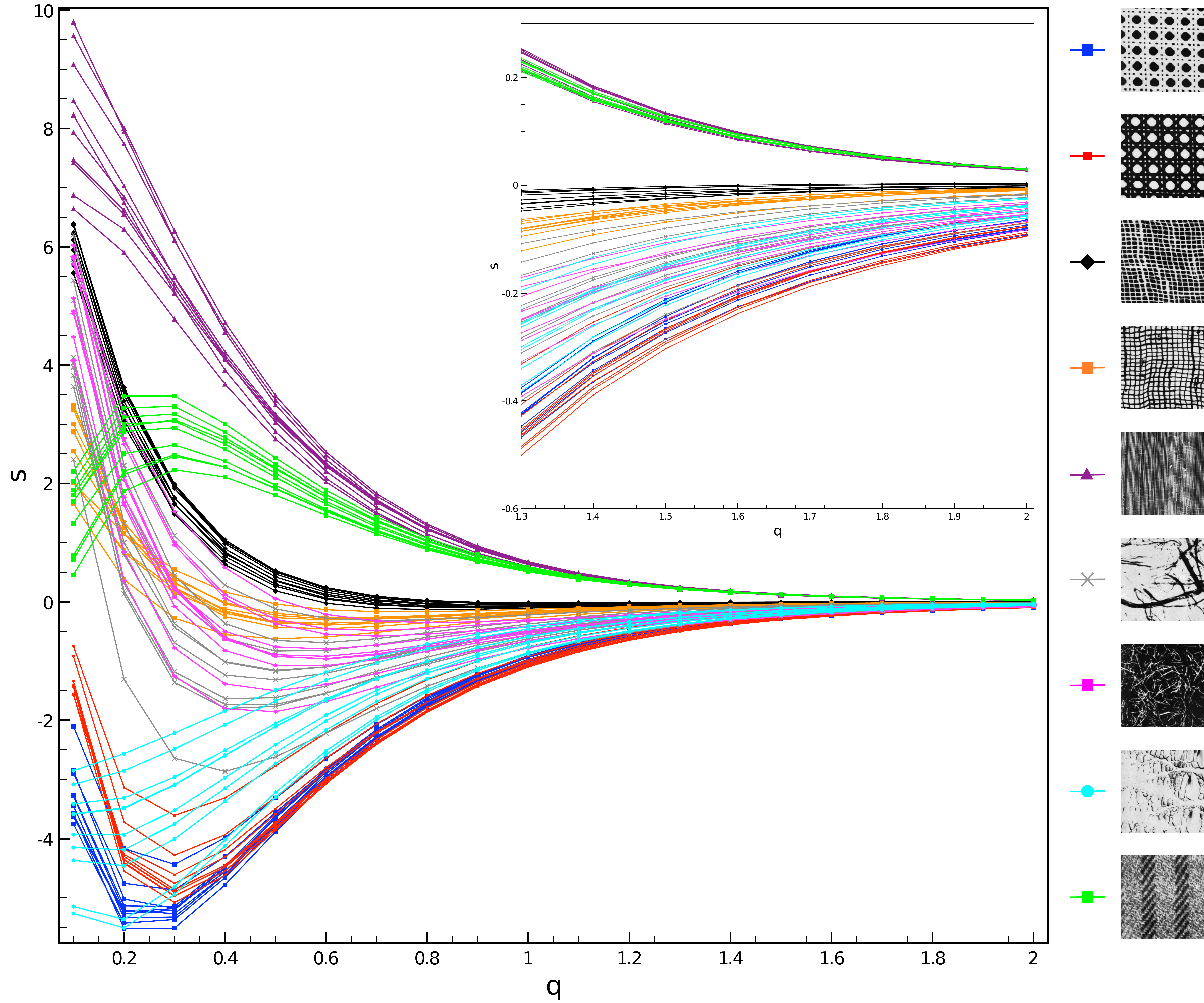}
\caption{%
Tsallis entropy versus $q$ for different patterns. Each color and symbol
combination represents a class of texture pattern.
}\label{fig:qxs}
\end{figure}

Since the nature of the $S_q$ curve is exponential, it is difficult to grasp the
differences between the patterns. In order to improve the visualization of
pattern behavior through the $S_q$ curve, we calculated the mean vector
$\vec{\mu}$, that is, the average $S_q$ curve of the 400 samples of the image database, and
plotted the difference of $\vec{S_q}$ and $\vec{\mu}$ for 10 patterns picked at
random, Figure~\ref{fig:qxs}. Notice that the first values of $q$ present the best
pattern discrimination, which agrees with the results of the experiments for a
single $q$, where a $q$ around $0.2$ performs best.

To use Tsallis entropy curve as a pattern recognition tool, we composed a feature vector $\vec{S_q}$ in the interval q=0.1:0.1:2 (\ie, $q$ from 0.1 to 2 in increments of 0.1). 
Using this feature vector of 20 elements, a classification rate of $73.75\%(\pm 6.49)$ was achieved.
We also constructed feature vectors using different intervals of values of $q$.
The classification results are shown in Table~\ref{tab:rangeofqs}.
These results corroborate the hypothesis that a multi-q approach can improve the power of the Tsallis entropy applied to pattern recognition.
The multi-q strategy using only 20 elements results in a gain of $41\%$ compared to best value of $q$ and a gain of $48.75\%$ compared to the BGS entropy.

\begin{table}[h]
\centering
\begin{tabular}{|l|c|c|}
\hline
\textbf{Range of $q$} & \textbf{\#Features} & \textbf{Classification Rate \%} \\\hline
\hline
0.2 & 1 & $32.75(\pm 6.17)$ \\\hline
0.5:0.5:2 & 5 & $52.75(\pm 6.27)$ \\\hline
0.2:0.2:2 & 10 & $67.00(\pm 6.43)$ \\\hline
0.1:0.1:2 & 20 & $73.75(\pm 6.21)$ \\\hline
0.05:0.05:2 & 40 & $75.75(\pm 6.54)$ \\\hline
0.01:0.01:2 & 200 & $77.25(\pm 6.49)$ \\\hline
0.005:0.005:2 & 400 & $77.50(\pm 6.01)$ \\\hline
0.001:0.001:2 & 2000 & $78.50(\pm 6.16)$\\\hline
\end{tabular}
\caption{Classification results for different sets of $q$.}
\label{tab:rangeofqs}
\end{table}

\begin{figure}[!hp]
\centering
\includegraphics[width=0.99\textwidth]{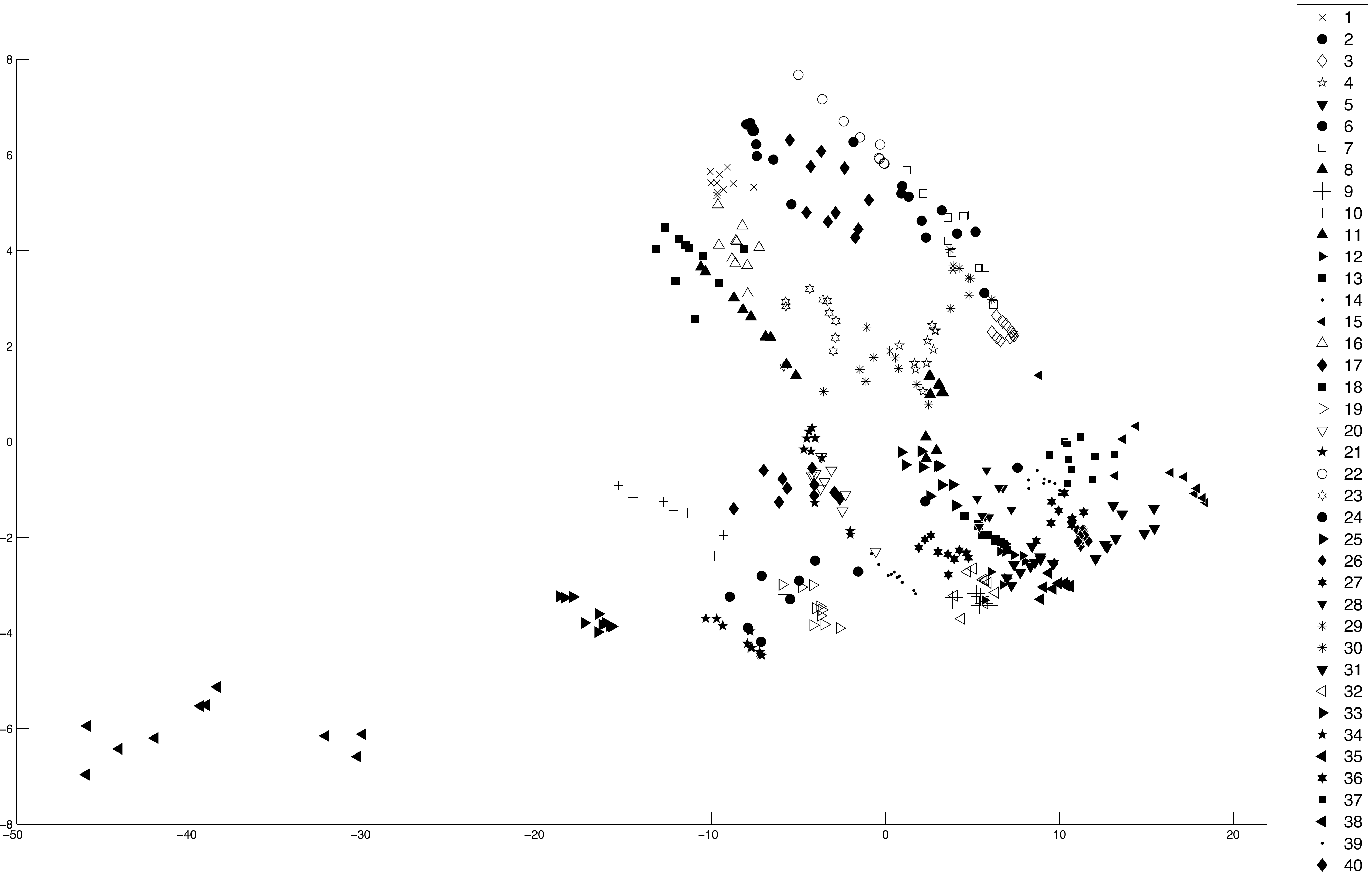}
\caption{To give a visual idea of the discriminating power of the Tsallis
entropy, a principal components analysis (PCA) was performed on Brodatz's 40
classes considering a 20-dimensional feature vector $(S_{0.1},..., S_2)$, and visualized as
a 2-dimensional scatter plot of the first two principal components.}
\label{fig:pca}
\end{figure}

To visualize the behavior of the texture classes, we use a Karhunen-Loeve transform (or principal components analysis, PCA).
This allows us to projecting the feature vectors onto a lower-dimensional space which
is easier to visualize, and where the variance is higher as possible.
The PCA  was applied to the feature vectors of the 400 samples (40 classes) of
the Brodatz dataset and a scatter plot was obtained, Figure~\ref{fig:pca}.
As we can see, the classes become organized in distinguished cluster, illustrating the power of classification of the multi-q approach. 

\subsection{Feature selection: enhancing the discriminanting power of the $S_q$ curve}

We have noticed that composing a feature vector with different $S_q$'s can
boost the discriminative power of the Tsallis entropy. Nevertheless, the
feature vector was composed by $q$ in the range $0.1,0.2,\dots 2$ and, as remarked in Figure~\ref{fig:classific:versus:q}, 
there are $q$ values in the interval that do not achieve the maximum
classification. Therefore, a question arises: 
is the information of the entire interval $0 < q \leq 2$ aiding to distinguish the image patterns? 

\begin{figure}[!h]
\centering
\footnotesize{(a)}\includegraphics[width=0.425\textwidth]{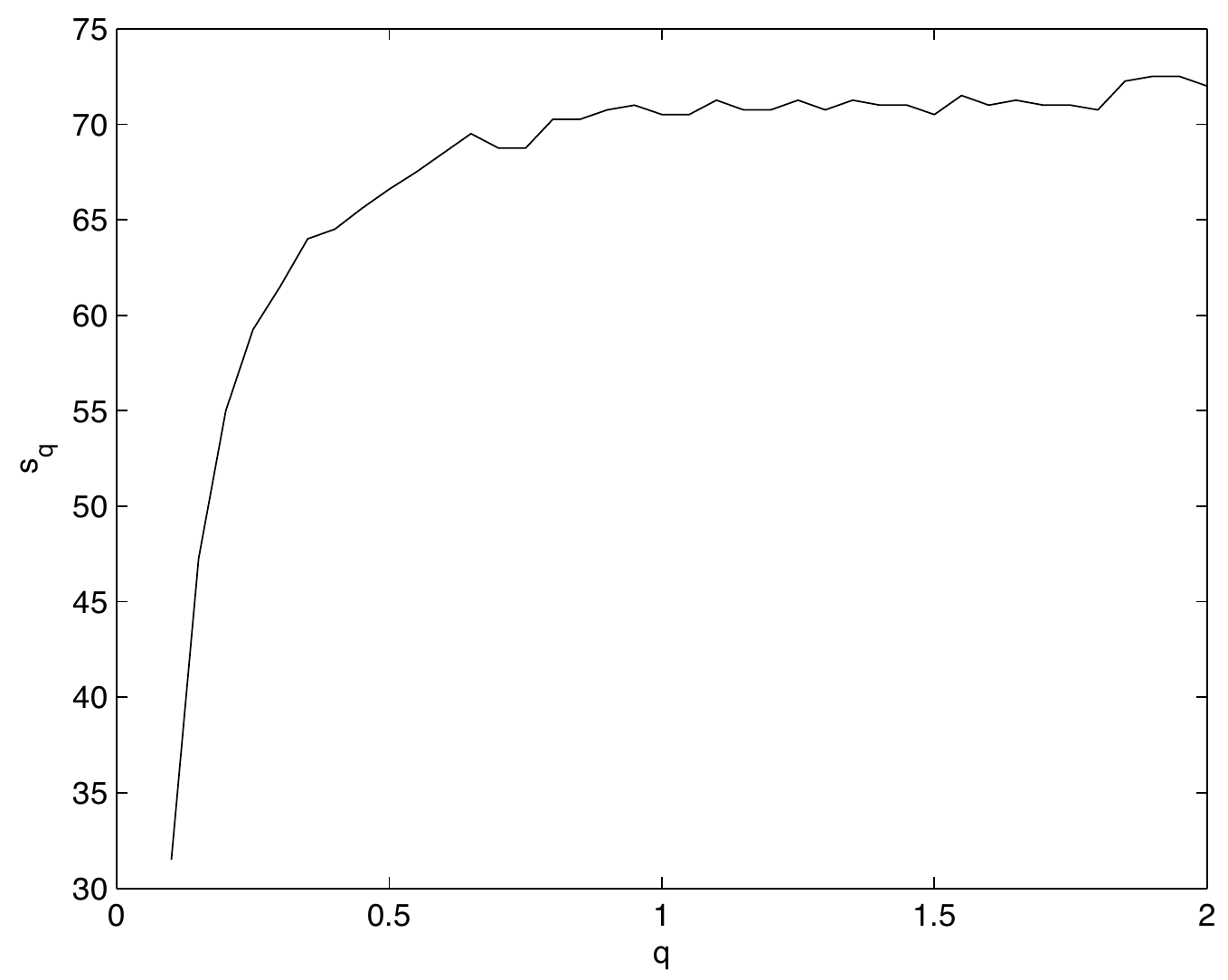}%
\footnotesize{(b)}\includegraphics[width=0.5\textwidth]{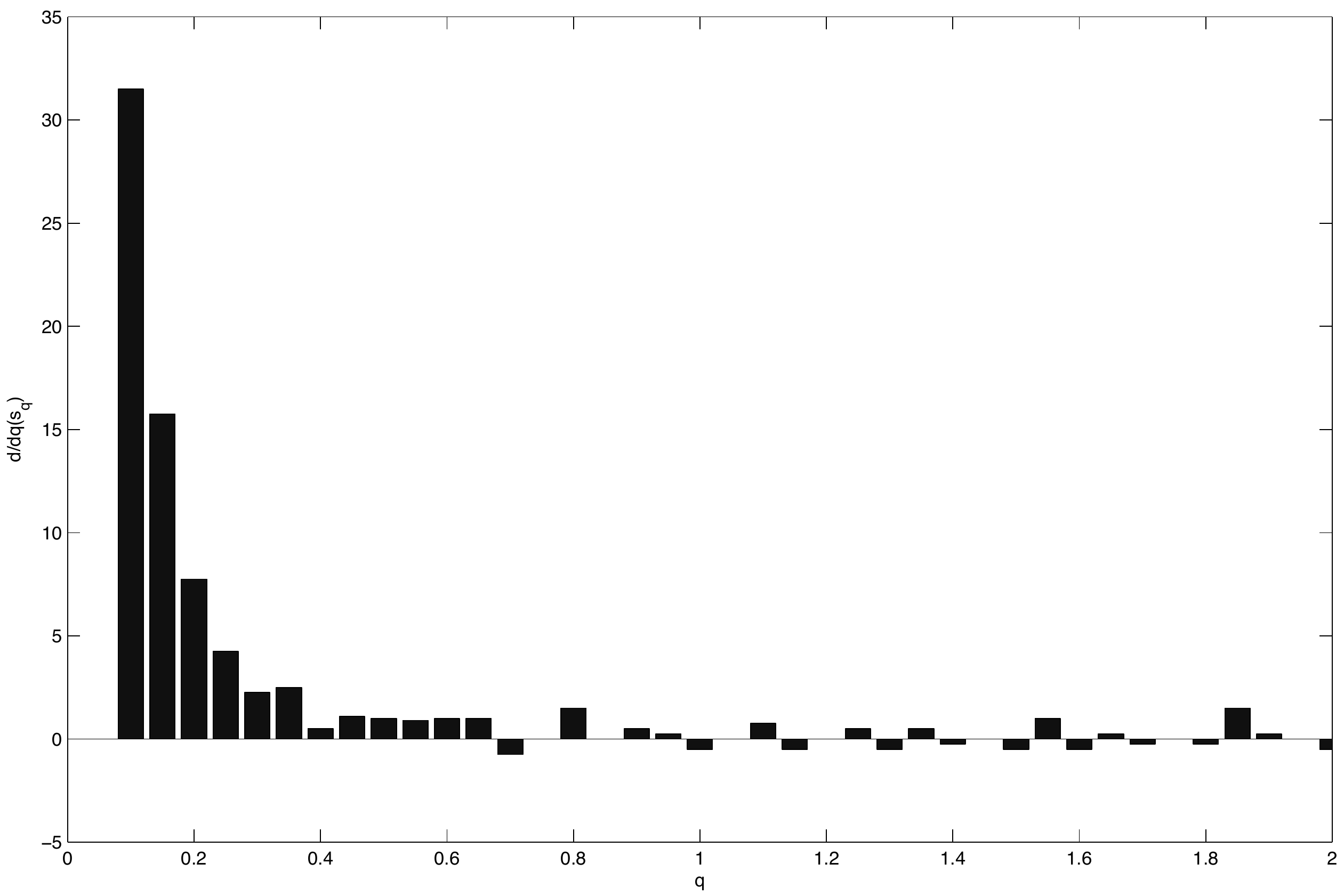}%
\caption{%
(a) Classification rate for feature vectors using the first $S_q$ elements: for
a value of $q$, the curve gives the classification when using
$\vec S_q = (S_{0.1},\dots,S_q)$ as a feature vector. The number of elements in
the feature vector increases for each element of x axis. 
(b) the derivative of the previous curve.
}\label{fig:feature_selection1}
\end{figure}

Feature selection is a technique used in multivariate statistics and in pattern
recognition that selects a subset of relevant features with the aim of improving
the classification rate and also its robustness.  There are several algorithms
for feature selection, the reader can find a feature selection survey in~\cite{Molina:etal:ICDM2002}. We
have used a very simple strategy in the present work to clarify the influence of the
different $q$ in the classification process. The main idea is use only the $S_q$
values that presents significant contribution to the classification rate. Figure~\ref{fig:feature_selection1}(a) plots the classification rate using the first
$S_q$'s, taking different quantities of these; \eg, for the first
datapoint of the
curve a single $S_q$ was used, $(S_{0.1})$, in the second datapoint two $S_q$'s,
$(S_{0.1}, S_{0.2})$, and at 
position $n$ on the $x$ axis, $n$ $S_q$'s. The curve shows
there are values of $S_q$ that improve the classification rate
but there are also values of $q$ that do not increase the classification rate or
even decrease it. To make the contribution of each $S_q$ easier to see, we
take the derivative of the curve of the Figure~\ref{fig:feature_selection1}(a),
shown in the Figure~\ref{fig:feature_selection1}(b). We performed feature selection
by picking the $q$'s whose values in the derivative curve are greater than $t$.

The feature selection reduces the number of features and increases the
classification power of the multiple $q$ entropy approach. The Table
\ref{tab:rangeofq} shows results of the feature selection over multi-$q$
approach for different range of $q$ at the interval 0 to 2. As can be observed
with just 4 elements the result is equivalent to a feature vector with size 20
without the feature selection (see Table \ref{tab:rangeofqs}) and with 27
elements, the feature selection approach overcome the performance of a 2000 size
feature vector (Table \ref{tab:rangeofqs}). The results demonstrates the feature
selection algorithm presents a optimal performance of the mult-$q$ approach. The
main reason of the performance increase is that the algorithm can select the
$S_q$ significant elements.  

The confusion matrices for different representative approaches to entropy-based
image classification investigated in this paper are shown in 3D
Figure~\ref{fig:feature:selection1}. An arbitrary entry $(x,y)$ of a confusion matrix
expresses how many patterns of class $x$ were labeled as class $y$, and this is
visualized as height $z$ in the figure. This visualizations gives insight into
the error of the classification and which class was most wrongly classified.

\begin{table}[h]
\centering
\begin{tabular}{|l|c|c|}
\hline
\textbf{Range of $q$} & \textbf{\#Features} & \textbf{Classification Rate \%} \\\hline
\hline
0.5:0.5:2 & 4 & 52.75 \\\hline
0.2:0.2:2 & 4 &73.75 \\\hline
0.1:0.1:2 & 6 &80 \\\hline
0.05:0.05:2 & 7 &80.25 \\\hline
0.01:0.01:2 & 21 &81.5 \\\hline
0.005:0.005:2 & 26 &81.75 \\\hline
0.001:0.001:2 & 27 &82 \\\hline
\end{tabular}
\caption{Classification results for feature selection from different step sizes of $q$}
\label{tab:rangeofq}
\end{table}


\begin{figure}[!h]
\centering
(a)\includegraphics[width=0.45\textwidth]{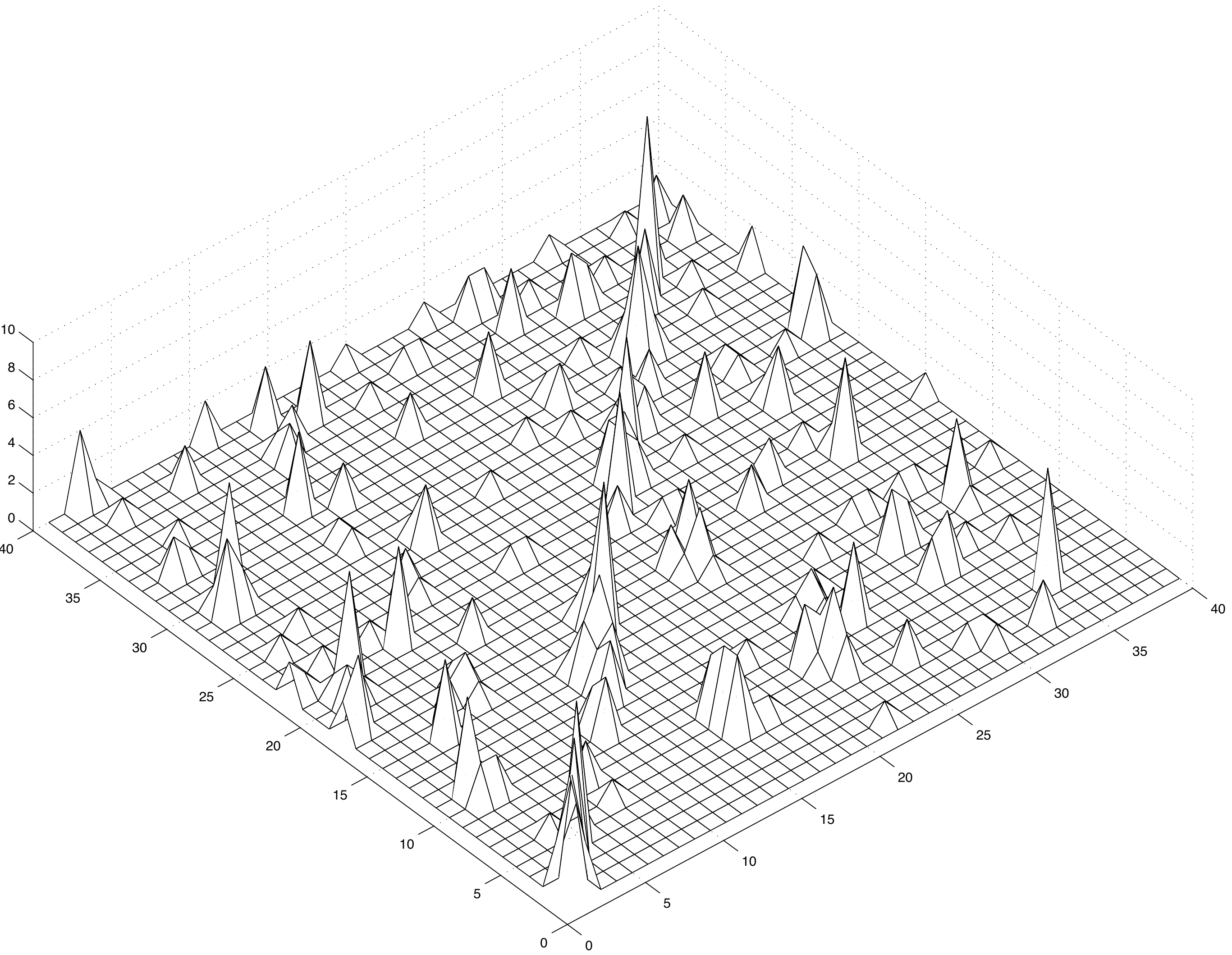}%
(b)\includegraphics[width=0.45\textwidth]{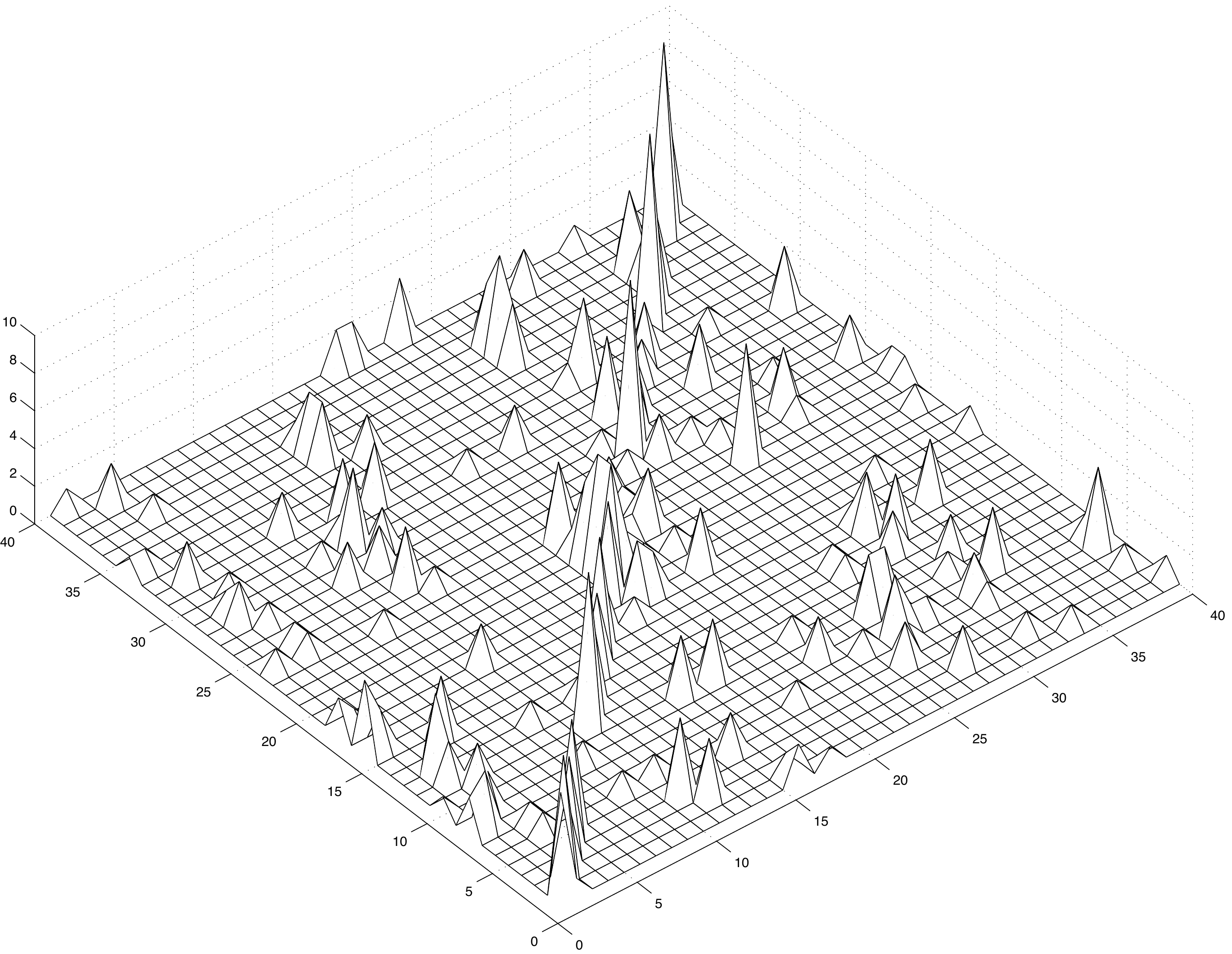} \\
(c)\includegraphics[width=0.45\textwidth]{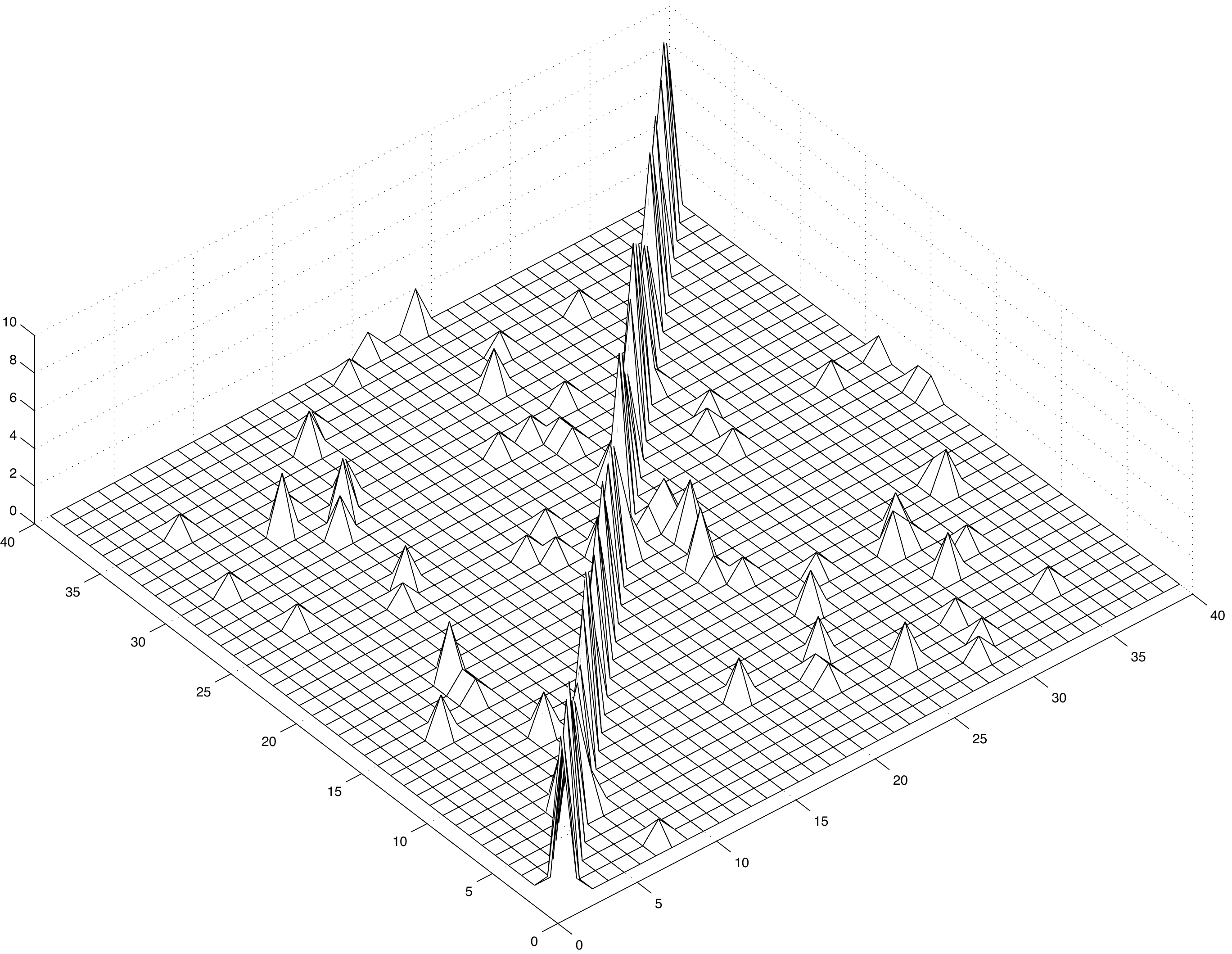}%
(d)\includegraphics[width=0.45\textwidth]{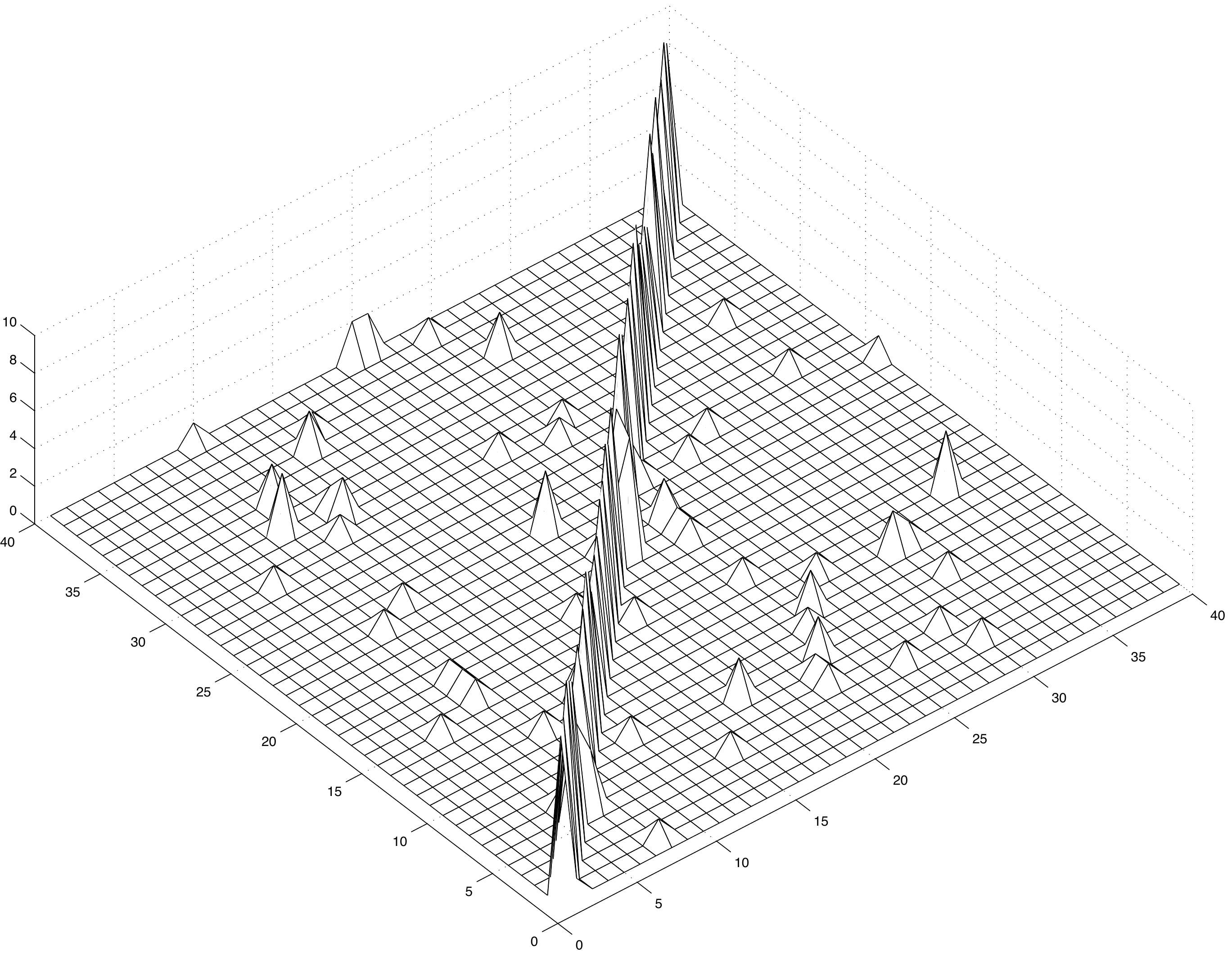}%
\caption{
Confusion matrices for (a) BGS entropy alone, (b) Tsallis entropy with $q=0.2$,
(c) multi-$q$ approach using the feature vector $\vec S_q = (S_{0.1},\dots,S_2)$,
and (d) feature vector constructed with the feature selection technique described in the
paper.}
\label{fig:confusion:matrix:3D}
\end{figure}

\section{Conclusion}\label{sec:conclusion}
In this paper we showed how the Tsallis entropy can be used in image pattern
classification with great advantage over the classic entropy.  The parametrized
Tsallis entropy enables larger-dimensional feature vectors using
different values of $q$, which yields vastly better performance than
using BGS entropy alone. This points to the fact that the Tsallis entropy for
different $q$ \emph{does} encode much more information from a given histogram than the BGS
entropy. In fact, one of the results show that as little as 4 values of $q$, together,
are enough to outperform the BGS entropy by about $3\times$.  Work to further analyze the
implications of these results within a deeper information-theoretic framework is
underway, shedding light into the usefullness of the Tsallis entropy for general
problems of pattern recognition. 

\section*{Acknowledgements}
R.F. acknowledges support from the UERJ visiting professor grant.
W.N.G. acknowledges support from FAPESP (2008/03253-9). 
F.J.P.L. were supported by the CNPq/MCT, Brazil and the Research Foundation of the State of Rio de Janeiro (FAPERJ, Brasil). 
O.M.B. acknowledges support from CNPq (Grant \#308449/2010-0 and \#473893/2010-0) and FAPESP (Grant \# 2011/01523-1). 


\end{document}